\documentclass[12pt]{article}
\usepackage[utf8]{inputenc}
\usepackage{authblk}
\usepackage{setspace}
\usepackage[margin=1.25in]{geometry}
\usepackage{graphicx}
\usepackage{subcaption}
\usepackage{amsmath}
\usepackage{lineno}
\usepackage{lipsum}
\usepackage[numbers,square,compress]{natbib}

\begin{document}
\title{Determination of Lattice QCD Equation of state at a finite chemical potential}
\author[1*]{Sabarnya Mitra}

\affil[1]{Centre for High Energy Physics, Indian Institute of Science, Bengaluru - 560012}

\newcommand{\Ob}{\mathcal{O}}
\newcommand{\Z}{\mathcal{Z}}

\onehalfspacing
\maketitle

\begin{abstract}

Exponential resummation to all orders in $\mu_B$ is a promising 
scheme which can capture contributions to all orders in $\mu_B$ by considering only the first few Taylor coefficients in the Taylor expansion of thermodynamic observables in lattice QCD. This approach however, gets affected by biased estimates of $n$-point correlation functions $D_n$ whose effects can significantly hinder probe of finite density QCD. We present an unbiased exponential resummation formalism, which can reproduce the Taylor series up to the desired order in $\mu_B$, besides retaining the original form of the all-order resummation estimate of QCD partition function. 
\end{abstract}

\section{Introduction}

The Quantum Chromodynamics (QCD) phase diagram\,\cite{Halasz:1998QCDph,Karsch:2001QCDph,Fukushima:2010QCDph} is an intriguing spectacle of QCD, offering deep insights about the different phases and properties of strongly interacting matter. The determination of QCD Equation of state (EoS) at finite temperature and density\,\cite{Allton:2002EoS,Bazavov:2009EoS,Borsanyi:2013EoS,HotQCD:2014EoS} is pivotal for investigating this phase diagram. One significant hurdle in knowing QCD EoS is the fermion sign problem in QCD partition function and the Taylor expansion method of thermodynamic observables is one of the numerous approaches\,\cite{Fodor:2001reweighting,deForcrand:2002analytic,Cristoforetti:2012thimble,Bollweg:2022Pade} adopted to circumvent this problem in lattice QCD. Due to computational and precisional difficulties regarding calculation of higher order Taylor coefficients, one turns towards exponential resummation. In this approach, one expresses the partition function in an exponential form, exponentiating the contributions of first $N$ baryon correlation functions and resumming to all orders in $\mu_B$. Despite this improvement, this formalism gets affected by stochastic bias, which can radically misdirect results and consequent implications. Although our work of cumulant expansion\,\cite{Mitra:2022prd} promises to eliminate this bias order-by-order, this however caused us losing the invaluable all-order resummation altogether. In this work, I will present the formalism of unbiased exponential resummation, which eliminates stochastic bias and replicates the Taylor series completely upto a desired order in $\mu_B$.  

\section{Formalism : A brief note}

Using a total of $\text{N}_{\text{R}}$ Gaussian random volume sources for every gauge configuration at a given temperature $T$, the exponential resummed estimate of the QCD partition function to $\Ob(\mu_B^N)$ resembles 

\begin{equation}
    \Z_N^{\text{R}} = \left \langle \text{Re}\left[\exp\left(\sum_{n=1}^N \mathbf{D}_n \left(\frac{\mu_B}{T}\right)^n \right) \right]\right  \rangle, \hspace{3mm} \text{where} \hspace{6mm}\mathbf{D}_n = \frac{1}{{\text{N}}_{\text{R}}} \sum_{r=1}^{\text{N}_{\text{R}}} D_n^{(r)}
    \label{eq:old resum}
\end{equation}
On expansion in orders of $\mu_B$, the above resummed estimate of $\Z$ in Eqn.\eqref{eq:old resum} gives rise to biased estimates\,\cite{Mitra:2022prd} of $\mathbf{D}_n$ in which the estimate of a given random source among all in the distribution, is raised to higher non-linear powers. In the unbiased estimate, all the estimates of the sample $D_n^{(r)}, 1 \leq r \leq N_R$ are raised to linear powers providing equal importance to all the random estimates. It is important to eliminate these biased estimates for obtaining correct results.
We present the formalism of unbiased exponential resummation in chemical potential ($\mu_B$) and cumulant ($X$) bases. In $\mu_B$ and $X$ bases, we have
 \begin{align}   
     \frac{\Delta P_{N}^{\text{u},\mu_B}}{T^4} &= \frac{1}{VT^3} \hspace{1mm}\ln \hspace{1mm} \mathcal{Z}_{N}^{\text{u},\mu_B} , \hspace{2mm}\mathcal{Z}_{N}^{\text{u},\mu_B} = \left \langle e^{A_N(\mu_B)} \right \rangle , \hspace{2mm} 
    A_N(\mu_B) = \sum_{n=1}^{\boldsymbol{N}} \left(\frac{\mu_B}{T}\right)^n\frac{\mathcal{C}_{n}}{n!} 
    \label{eq:mu basis}\\ 
     \frac{\Delta P_{N}^{\text{u},X}}{T^4} &= \frac{1}{VT^3} \hspace{1mm}\ln \hspace{1mm} \mathcal{Z}_{N}^{\text{u},X} , \hspace{2mm}\mathcal{Z}_{N}^{\text{u},X} = \left \langle e^{Y_N^M(X)} \right \rangle , \hspace{2mm} 
    Y_N^M(X) = \sum_{n=1}^{\boldsymbol{M}} \frac{\mathcal{L}_{n}(X_N)}{n!}
    \label{eq:X basis}
      \end{align}
The detailed expressions of $C_n$ and $\mathcal{L}_n$ in above Eqns.\eqref{eq:mu basis} and \eqref{eq:X basis} are given in Ref.\,\cite{Mitra:2022prd}. Our results have been obtained using $N=2,4$ and $M=4$.


\section{Results}

\begin{figure}[!htb]
  \centering
   \includegraphics[width=0.325\textwidth]{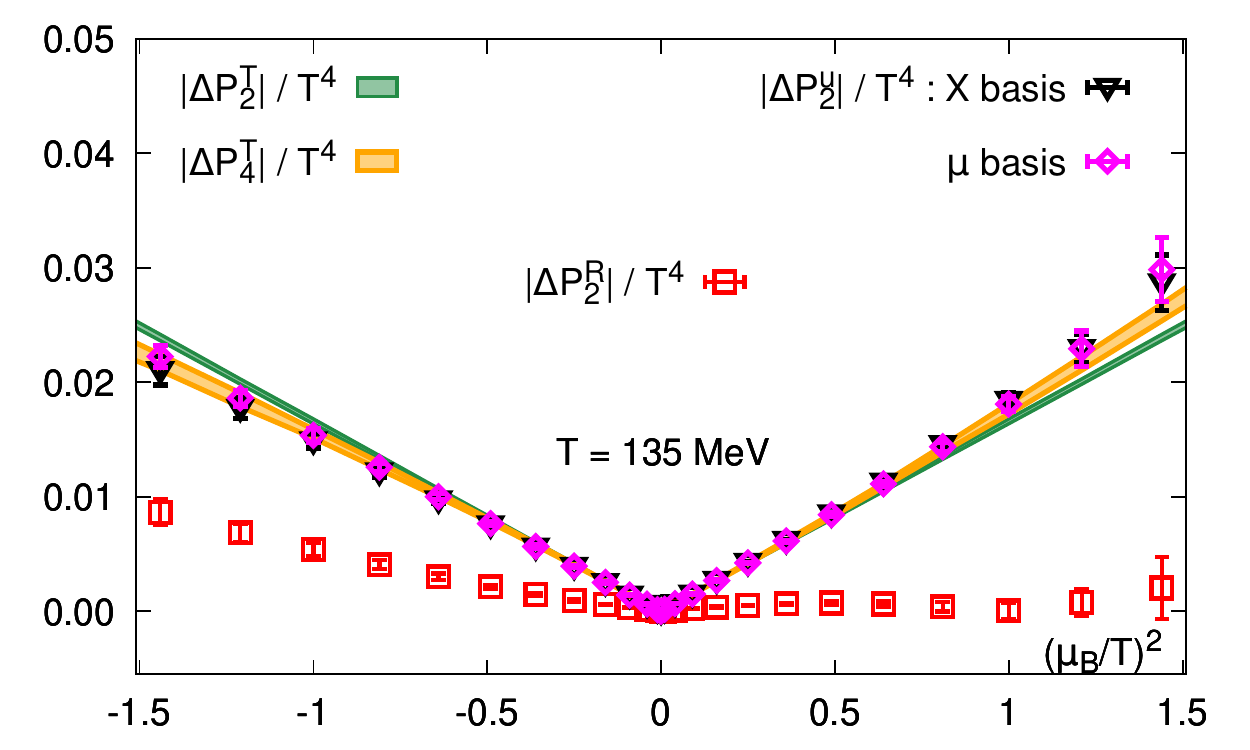} 
    \includegraphics[width=0.325\textwidth]{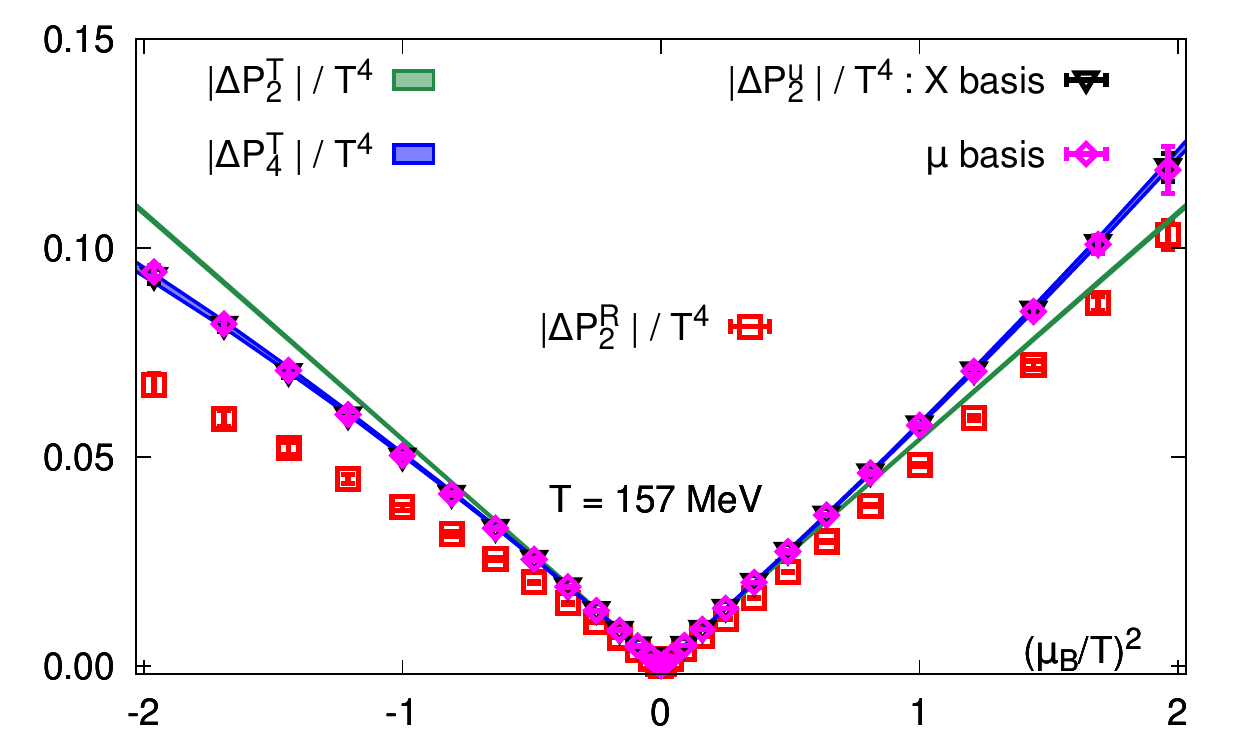} 
    \includegraphics[width=0.325\textwidth]{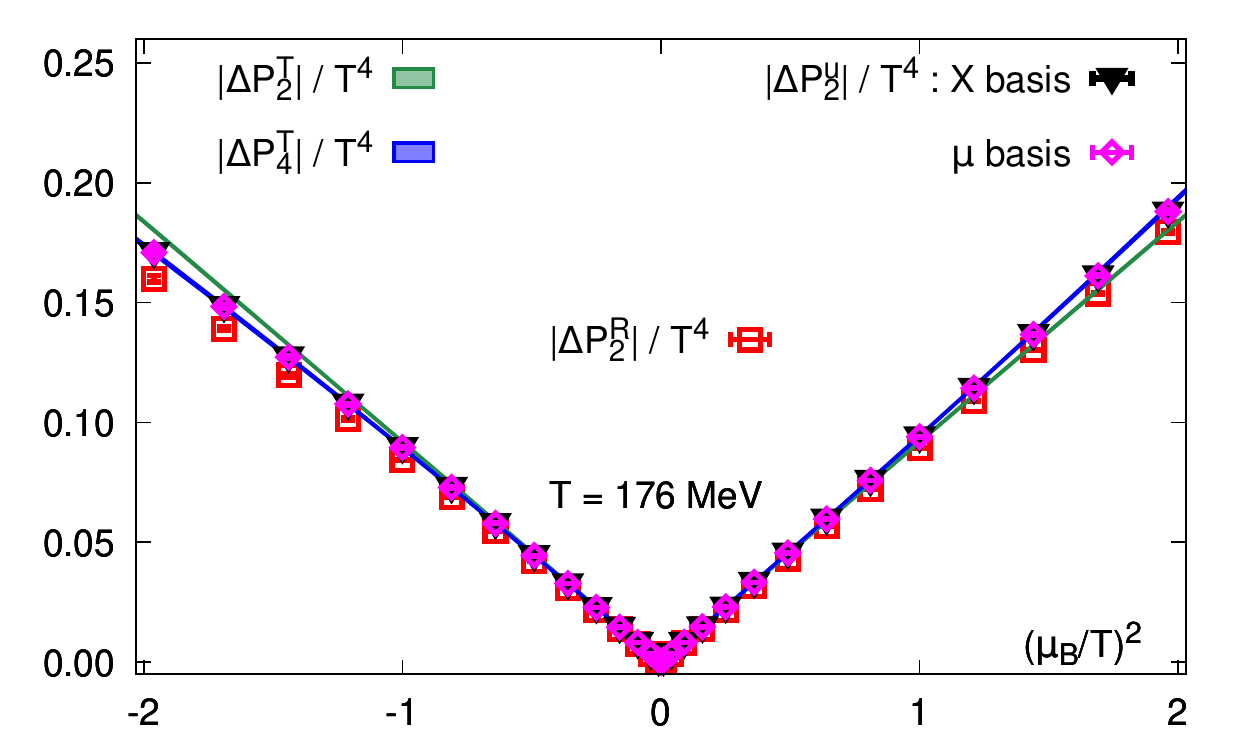} \\
     \includegraphics[width=0.325\textwidth]{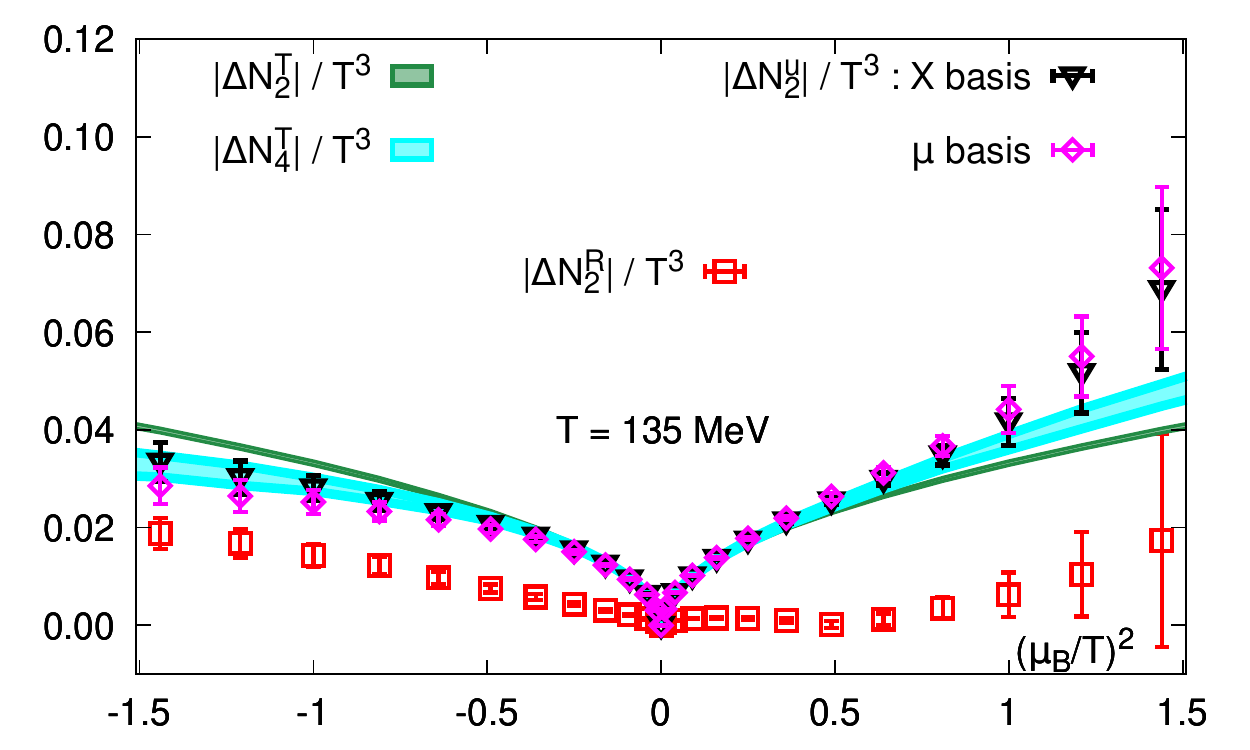} 
     \includegraphics[width=0.325\textwidth]{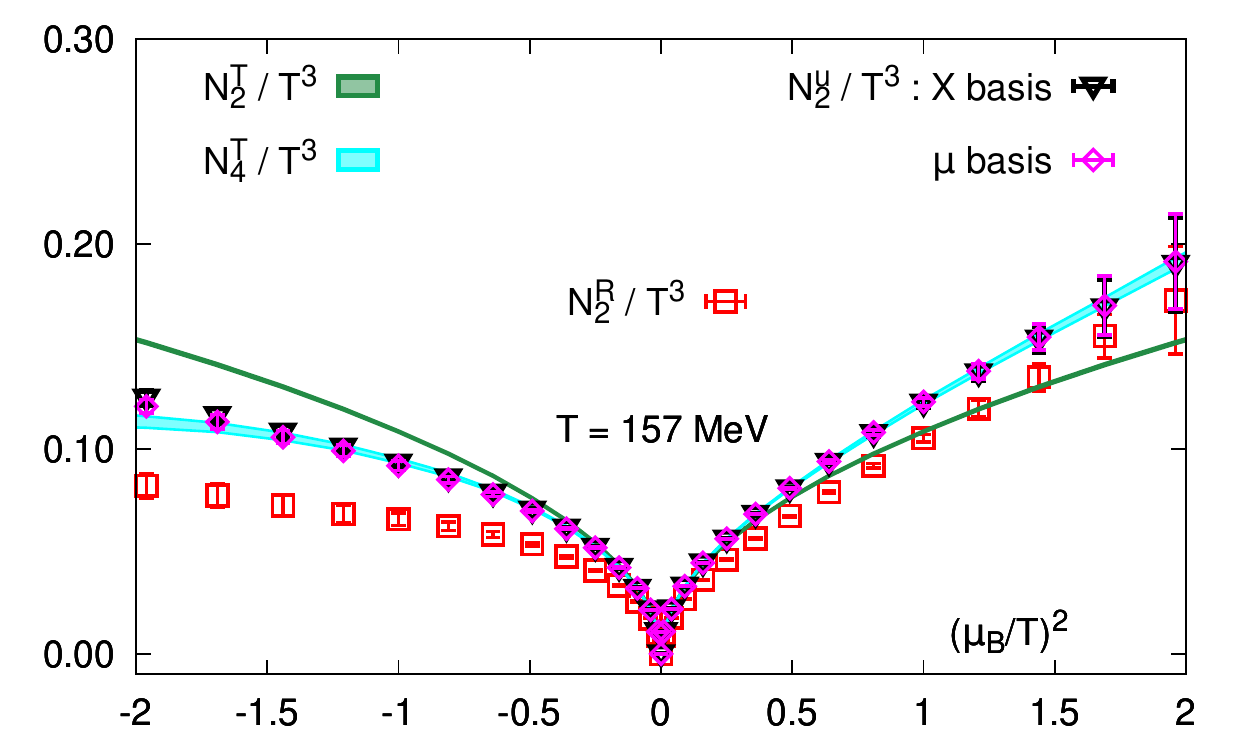}
    \includegraphics[width=0.325\textwidth]{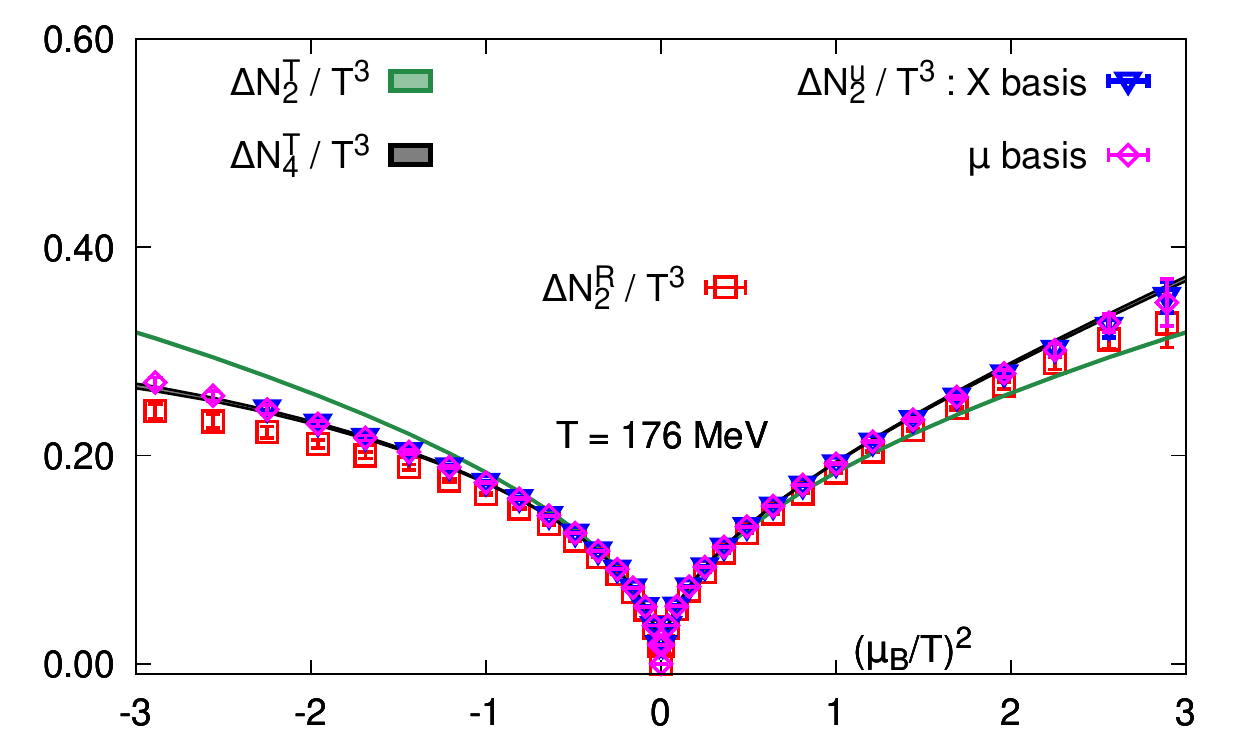}
  \caption{$\Delta \text{P}_2/T^4$ (top row) and $\text{N}_2/T^3$ (bottom row) plots in $(\mu_B/T)^2$ for \\$135$ (left column), $157$ (middle column) and $176$ (right column) MeV}
  \label{fig:all T}
\end{figure}

\begin{figure}[!tb]
     \centering
     \includegraphics[width=0.40\textwidth]{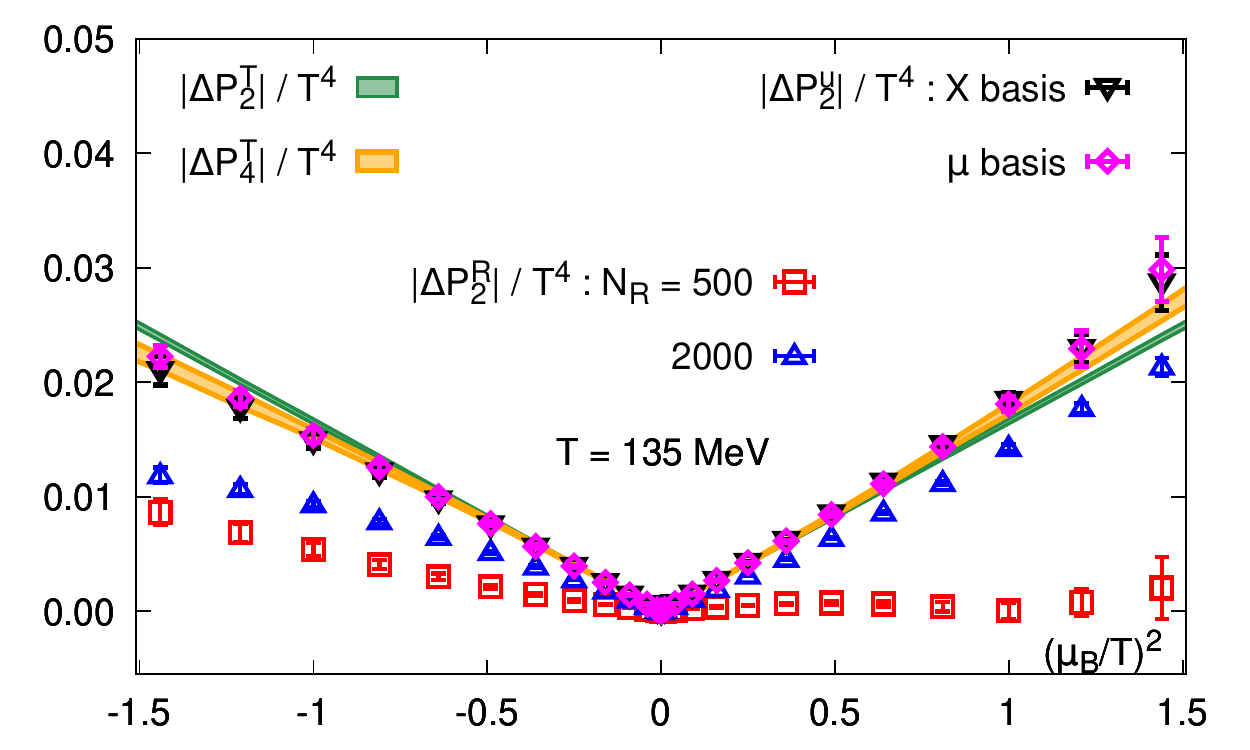} 
     \includegraphics[width=0.40\textwidth]{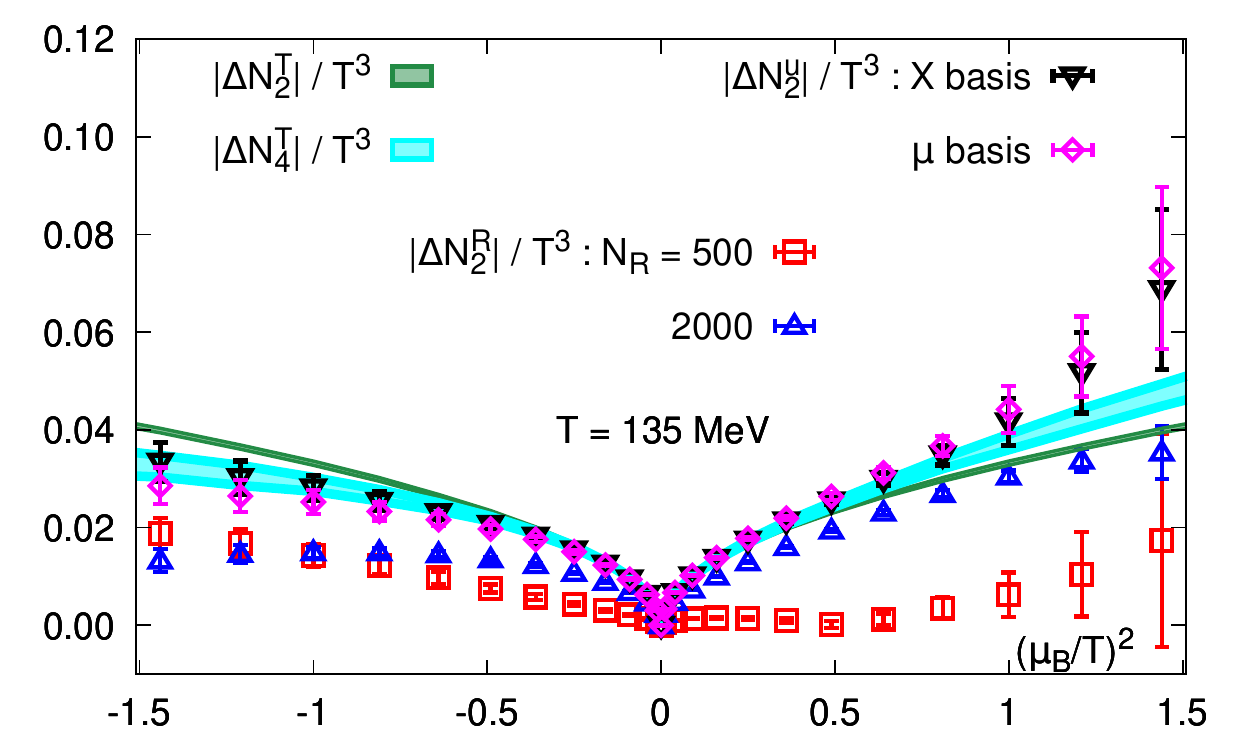} \\
     \includegraphics[width=0.40\textwidth]{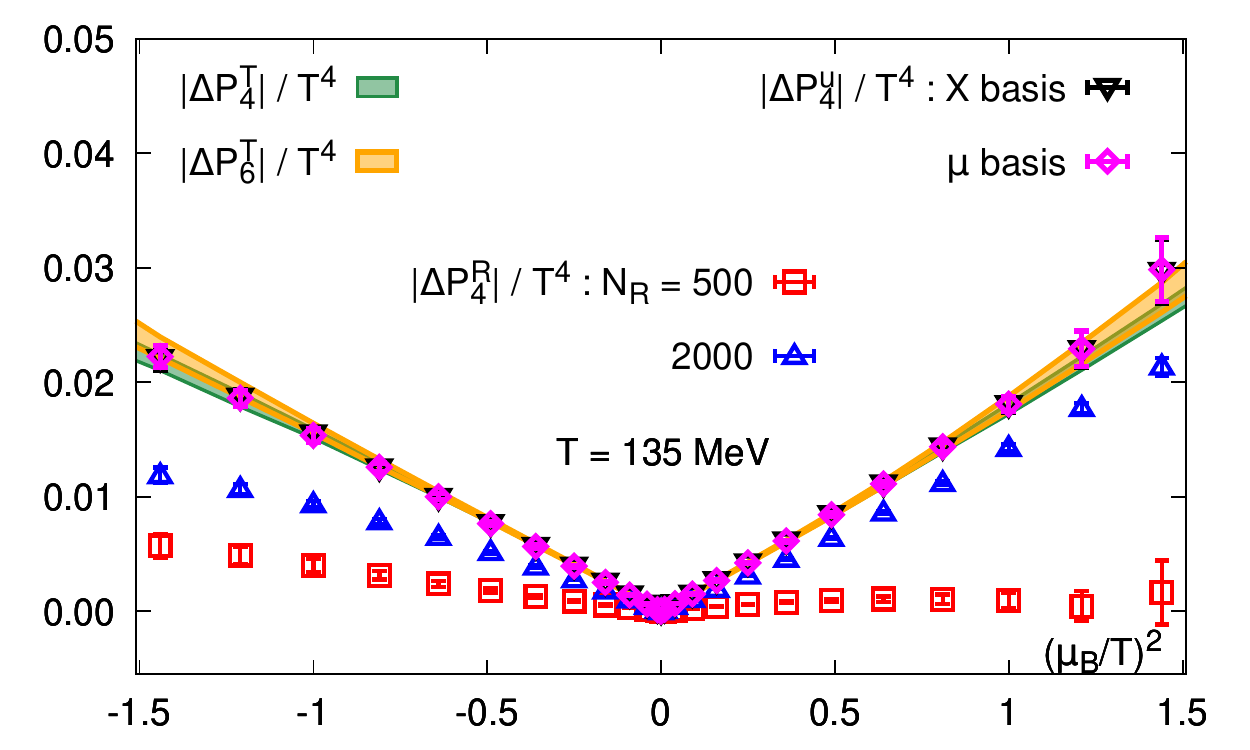} 
     \includegraphics[width=0.40\textwidth]{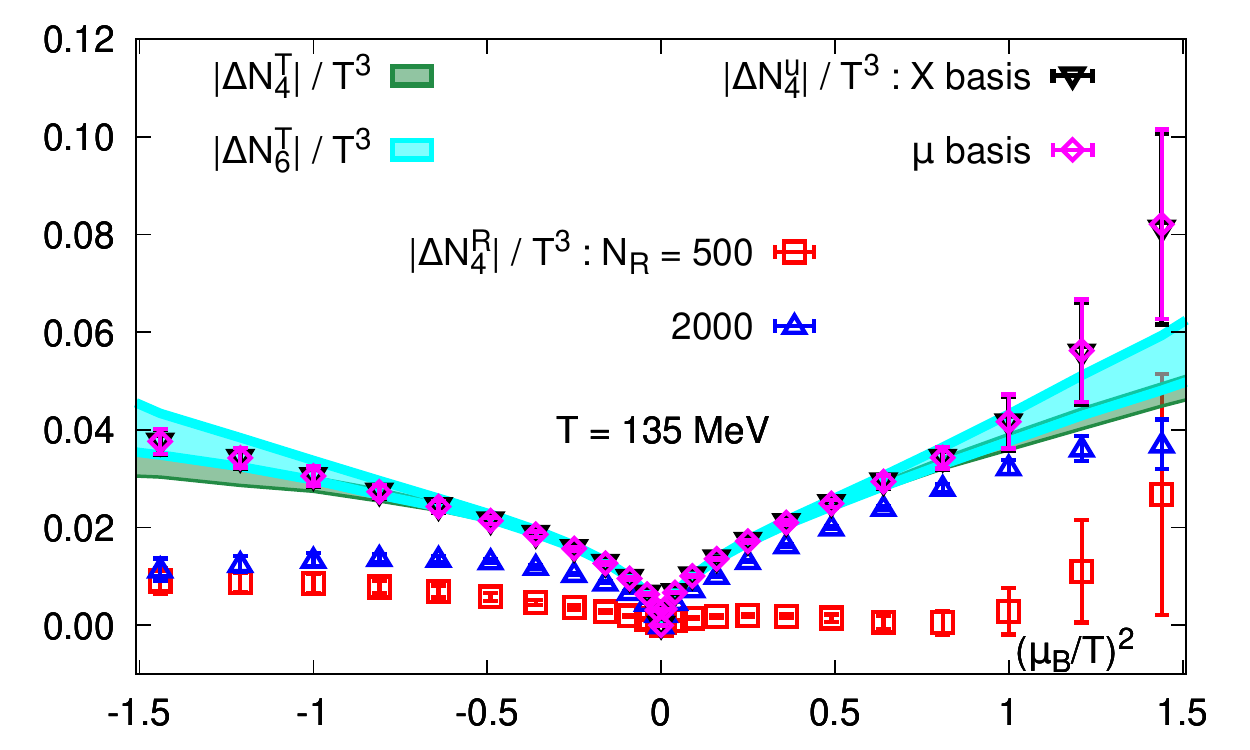} \\
    \caption{$\Delta \text{P}_{2,4}/T^4$ (left) and $\text{N}_{2,4}/T^3$ (right) plots in both bases for $T = 135$ MeV}
       \label{fig:2000}
        \end{figure}

This analysis has been performed in $2+1$ flavor QCD using Highly Improved Staggered Quarks action and a Symanzik-improved gauge action with $100$K gauge configurations. Fig.\ref{fig:all T} indicates that the stochastic bias is highly dominant at $135$ MeV\,\cite{Mitra:2023prl}, and progressively reduces with increasing temperatures at $157$ and $176$ MeV\,\cite{Mitra:2023prd}. They signify the three phases of the phase diagram. This is indicated by the near-alignment of the red points $\Delta \text{P}^{\text{R}}$, $\text{N}^{\text{R}}$ with the Taylor series bands $\Delta \text{P}^{\text{T}}$, $\text{N}^{\text{T}}$ in Fig.\ref{fig:all T}. Despite being very pronounced at $135$ MeV, Fig.\ref{fig:all T} demonstrates this new formalism not only eliminates bias upto $\Ob(\mu_B^2)$, but also very efficiently captures the unbiased contributions and agrees well with the fourth order Taylor results of excess pressure and number density. Although stochastic bias is significantly reduced by using old resummation formula with $2000$ random sources for $D_1$, the new formalism with just $500$ sources for all $D_n$ shown in Fig.\ref{fig:2000} provides more improved results and manages to capture higher order Taylor contributions over the former. 

\section{Conclusion and Outlook}

In this work, we have presented the new formalism of unbiased exponential resummation which successfully eliminates the much dreaded stochastic bias arising from the old exponential resummation formula to a given order in $\mu_B$. This new formalism, thus not only retrieve the structure of exponential resummation but also manages to capture genuine higher order Taylor contributions. Despite producing unbiased estimates upto a finite order in $\mu_B$, it implies that in the limit of infinite number of cumulants in $X$ basis or unbiased corrections to all orders in $\mu_B$ in $\mu_B$ basis, one can achieve unbiased exponential resummation to all orders in $\mu_B$ which is identical to the true infinite Taylor series. This reiterates the importance of obviating stochastic bias and following unbiased exponential resummation in obtaining genuine interpretations and implications regarding finite-density QCD and the mysteriously beautiful phase diagram of QCD.    

\section*{Acknowledgements}

I sincerely acknowledge Prasad Hegde and Frithjof Karsch for useful discussions. The computations have been performed in the cluster of Bielefeld University Germany, for which I thank the Bielefeld HPC.NRW team.

\end{document}